\documentstyle[12pt]{article}

\title{A non-perturbative approach
 to the Coleman-Weinberg
mechanism in massless scalar QED}

\author{A. P. C. Malbouisson \thanks{On leave
of absence from Centro Brasileiro de Pesquisas 
F\'isicas - CBPF} \\
Centre de Physique th\'eorique, CNRS UPR14 \\
Ecole Polytechnique \\
F-91128 Palaiseau Cedex \\
FRANCE \\
F. S. Nogueira \\
Centro Brasileiro de Pesquisas F\'isicas - CBPF \\
Rua Dr. Xavier Sigaud 150, Rio de Janeiro, RJ 22290-180 \\
BRAZIL \\
N. F. Svaiter \\
Centro Brasileiro de Pesquisas F\'isicas - CBPF \\
Rua Dr. Xavier Sigaud 150, Rio de Janeiro, RJ 22290-180 \\
BRAZIL}

\begin{document}

\maketitle

\newpage

\begin{abstract}
We rederive non-perturbatively the Coleman-Weinberg expression
for the effective potential for massless scalar QED. Our result is
not restricted to small values of the coupling constants. This
shows that the Coleman-Weinberg result can be established beyond
the range of validity of perturbation theory. Also, we derive it in
a manifestly renormalization group invariant way. It is shown
that with the derivation given no Landau ghost singularity
arises.
The finite temperature case is discussed.
\end{abstract}

\newpage

Ideas on situations which a system has a set of degenerate ground
states,
related by continuous symmetry transformations (spontaneous symmetry
breaking) originated in solid state physics
in the context of superconductivity
theory \cite{Anderson}. The Higgs phenomenon \cite{Higgs} appears as
 a
mechanism to explain spontaneous symmetry breaking. It postulates the
existence of a self-interacting scalar field (the Higgs field) which
produces massive vector fields in a locally gauge invariant
Lagrangean.
The Higgs "particle" results from the quantization of a model so
defined. For instance, in superconductivity the Higgs particle comes
from the breaking of a $U(1)$ symmetry and are the quanta of the
elementary excitations due to oscilations of the electron gas,
colective
modes of vibration, quasi-particles called plasmons. The plasmon mode
 is
associated to the generation of a massive degree of freedom to the
photon. This is the origin of the well known Meissner effect.

In particle physics this phenomenon is associated to the so called
standard model. The Higgs mechanism was adapted to particle physics
 in the context of the electroweak phenomenology
associated to the spontaneous breaking of a $SU(2){\times}U(1)$ 
symmetry \cite{Weinberg}.
According to the standard model, the
Higgs mechanism is responsible for the masses of the weak interaction
vector bosons, the $W^{\pm}$ and $Z^{0}$. However, the Higgs particle
has never been detected experimentally. The main
experimental evidence that the
Higgs mechanism in fact works is still associated to
superconductivity.
The search for the Higgs particle is presently one of the major goals
of experimental high energy physics.

In the Higgs model an imaginary mass is attributed to a charged
scalar field
which is self-interacting and is minimally coupled to a gauge field.
Then,
symmetry breaking is manifest already at the tree level. However,
even if the
mass of the scalar field is zero, we still have spontaneous symmetry
breaking induced by radiative corrections.  This is the
Coleman-Weinberg mechanism \cite{Coleman}.

The Coleman-Weinberg mechanism and its cosmological consequences was
extensively studied in the past decade
\cite{Guth,Sher,Witten,Billoire,Linde}. It was argued that in
theories of
grand unification the value of the coupling constant may become
large at
temperatures typically ${\sim}4{\times}10^{6}Gev$
\cite{Sher,Billoire,Linde}. In this regime
non-perturbative effects may
become important in the study of the phase transition. Therefore, it
will be of great interest to obtain the Coleman-Weinberg theory in a
non-perturbative scheme. We will show that this is possible at least
in the Abelian case.
In this paper we rederive the
Coleman-Weinberg theory  non-perturbatively in the case of massless
scalar QED.

Let us consider the Euclidean action for scalar QED in
$D=4$:

\begin{equation}
S={\int}d^{4}x\left[\frac{1}{4}F_{\mu\nu}F_{\mu\nu}+(D_{\mu}\phi)
^{\dag}
(D_{\mu}\phi)+m^{2}\phi^{\dag}\phi +\frac{\lambda}{3!}
(\phi^{\dag}\phi)^{2}
\right],
\end{equation}
where $F_{\mu\nu}=\partial_{\mu}A_{\nu}-\partial_{\nu}A_{\mu}$ and
$D_{\mu}=\partial_{\mu}-ieA_{\mu}$.

The vacuum-vacuum amplitude is given by the functional integral
representation:

\begin{equation}
Z=\int\prod_{x\mu}[d\phi^{\dag}(x)d\phi(x)dA_{\mu}(x)]\delta
[\partial_{\mu}
A_{\mu}]\det(-\Box)e^{-S}.
\end{equation}
Note in the functional measure the delta function
$\delta[\partial_{\mu}A_{\mu}]$ inforcing the Lorentz gauge
condition. The
factor $\det(-\Box)$ in the functional measure is the Faddeev-Popov
determinant.

It is convenient to reparametrize the theory in the unitary gauge.
This amounts in redefine the fields as

\begin{equation}
\phi=\frac{1}{\sqrt{2}}{\rho}e^{i\theta},
\end{equation}

\begin{equation}
B_{\mu}=A_{\mu}+\frac{1}{e}\partial_{\mu}\theta.
\end{equation}
The action becomes

\begin{equation}
S={\int}d^{4}x\left[\frac{1}{4}F'_{\mu\nu}F'_{\mu\nu}+\frac{1}{2}
\partial_{\mu}\rho\partial_{\mu}\rho+\frac{1}{2}e^{2}\rho^{2}
B_{\mu}B_{\mu}
+\frac{1}{2}m^{2}\rho^{2}+\frac{\lambda}{4!}\rho^{4}\right],
\end{equation}
where we have defined the new field strenght
$F'_{\mu\nu}=\partial_{\mu}B_{\nu}-\partial_{\nu}B_{\mu}$.

In terms of this new parametrization the functional integral is
written as

\begin{equation}
Z=\int\prod_{x\mu}[d\rho(x)\rho(x)dB_{\mu}(x)d\theta(x)]\delta
[\partial_{\mu}
B_{\mu}-\frac{1}{e}\Box\theta]\det(-\Box)e^{-S}.
\end{equation}
Since the action is independent of the field $\theta(x)$ it is
straightforward
to integrate out the gauge freedom:

\begin{equation}
\int[d\theta(x)]\delta[\partial_{\mu}B_{\mu}-\frac{1}{e}\Box\theta]=
\frac{\det(e)}{\det(-\Box)}=\frac{\prod_{x}e}{\det(-\Box)}.
\end{equation}
From Eq.(7) is readily seen that the Faddeev-Popov determinant is
cancelled out
from the functional measure.

Integration over the vector field $B_{\mu}$ yields

\begin{equation}
Z=\int\prod_{x}[d\rho(x)]e^{-S_{eff}[\rho]}
\end{equation}
where the effective action $S_{eff}[\rho]$ is given by

\begin{eqnarray}
S_{eff}[\rho] & = & \frac{1}{2}\ln\det[\delta_{\mu\nu}(-\Box+
e^{2}\rho^{2})+
\partial_{\mu}\partial_{\nu}]-\frac{1}{2}\delta^{4}(0){\int}d^{4}
x{\ln}
e^{2}\rho^{2} \nonumber \\
&   & +{\int}d^{4}x\left[\frac{1}{2}\rho(-\Box+m^{2})\rho+
\frac{\lambda}{4!}
\rho^{4}\right].
\end{eqnarray}
The above expression is exact. The factor
$-(1/2)\delta^{4}(0){\int}d^{4}x{\ln}e^{2}\rho^{2}$ arises from the
exponentiation of the factor $\prod_{x}e\rho(x)$ of the functional
measure.

The classical equation of motion is given by

\begin{equation}
\label{10}
\frac{{\delta}S_{eff}[\rho_{c}]}{\delta\rho(x)}=0.
\end{equation}
By considering a solution $\rho_{c}(x)=<\rho>=const.$, the
functional differentiation above becomes an ordinary differentiation
 and we obtain the
 following
equation:

\begin{equation}
{\int}d^{4}x\left(m^{2}<\rho>+\frac{\lambda}{3!}<\rho>^{3}-
\frac{\delta^{4}(0)}{<\rho>}\right)
+e^{2}<\rho>TrD_{\mu\nu}
(x-x')=0,
\end{equation}
where $D_{\mu\nu}(x-x')$ is the propagator of the massive vector
field with
mass $e^{2}<\rho>^{2}$.

Now, we have

\begin{equation}
TrD_{\mu\nu}(x-x')={\int}d^{4}x\left[3{\int}\frac{d^{4}p}
{(2\pi)^{4}}\frac{1}
{p^{2}+e^{2}
<\rho>^{2}}+\frac{\delta^{4}(0)}{e^{2}<\rho>^{2}}\right].
\end{equation}
Substituting (12) into (11) one obtains

\begin{equation}
{\int}d^{4}x<\rho>\left(m^{2}+\frac{\lambda}{6}
<\rho>^{2}+3e^{2}\int
\frac{d^{4}p}{
(2\pi)^{4}}\frac{1}{p^{2}+e^{2}<\rho>^{2}}\right)=0
\end{equation}
Note that the divergent factor $\delta^{4}(0)$ has been
cancelled out. Eq.
(13) gives us two possible solutions:

\begin{equation}
<\rho>=0,
\end{equation}
and
\begin{equation}
\label{15}
<\rho>^{2}=-\frac{6m^{2}}{\lambda}-\frac{18e^{2}}{\lambda}\int
\frac{d^{4}p}{
(2\pi)^{4}}\frac{1}{p^{2}+e^{2}<\rho>^{2}}.
\end{equation}
It is worth to emphasize the self-consistent character of Eq.(15). By
defining the parameters $M^{2}=e^{2}<\rho>^{2}$,
$M_{0}^{2}=-6m^{2}/\lambda$,
and $g=36e^{4}/\lambda$ we can rewrite (15) in the form

\begin{equation}
M^{2}=M_{0}^{2}-\frac{g}{2}\int\frac{d^{4}p}{(2\pi)^{4}}\frac{1}
{p^{2}+M^{2}}.
\end{equation}
Eq.(16) corresponds to the superdaisy resummation \cite{Dolan} for a
$\phi^{4}$ theory with negative coupling $-g$. The role of the mass
is
played by the parameter $M_{0}^{2}$.

By considering a $x$-independent background field $\overline{\rho}$
we
obtain the following expression for the effective potential:

\begin{equation}
V(\overline{\rho})=\frac{3}{2}\int\frac{d^{4}p}{(2\pi)^{4}}\ln
(p^{2}+e^{2}
\overline{\rho}^{2})+\frac{1}{2}m^{2}\overline{\rho}^{2}+
\frac{\lambda}{24}
\overline{\rho}^{4}.
\end{equation}
Note again the cancellament of the divergent factor $\delta^{4}(0)$.

Evaluating the integral in (15) by using a cutoff $\Lambda$ we obtain

\begin{eqnarray}
<\rho>^{2}\left(\lambda+\frac{9e^{4}}{8\pi^{2}}\ln\frac{\mu^{2}}
{\Lambda^{2}}
\right) & = & -6m^{2}-\frac{9e^{4}}{8\pi^{2}}\Lambda^{2} \nonumber \\
 &   & -\frac{9e^{4}}{8\pi^{2}}<\rho>^{2}\ln\frac{e^{2}<\rho>^{2}}{
\mu^{2}},
\end{eqnarray}
where $\mu$ is an arbitrary renormalization scale. From Eq.(18) we
define the renormalized quantities:

\begin{equation}
m_{R}^{2}=m^{2}+\frac{3e^{2}}{16\pi^{2}}\Lambda^{2},
\end{equation}

\begin{equation}
\lambda_{R}=\lambda+\frac{9e^{4}}{8\pi^{2}}\ln\frac{\mu^{2}}
{\Lambda^{2}}.
\end{equation}
Eq.(18) becomes

\begin{equation}
\label{21}
\lambda_{R}<\rho>^{2}=-6m_{R}^{2}-\frac{9e^{4}}{8\pi^{2}}<\rho>^{2}
\ln\frac{
e^{2}<\rho>^{2}}{\mu^{2}}.
\end{equation}
Also, by evaluating the integral in (17) using a cutoff
$\Lambda$ one obtains

\begin{equation}
\label{22}
V(\overline{\rho})=\frac{m_{R}^{2}}{2}\overline{\rho}^{2}+
\frac{\lambda_{R}}{
24}\overline{\rho}^{4}+\frac{3e^{4}}{64\pi^{2}}\overline{\rho}^{4}
\ln\frac{
e^{2}\overline{\rho}^{2}}{\mu^{2}}-\frac{3e^{4}}{128\pi^{2}}
\overline{\rho}^{
4}.
\end{equation}
Eq.(\ref{21}) together with Eq.(14) gives us the position of the
stationary points of the effective potential, Eq.(\ref{22}). If
$m_{R}^{2}>0$ the physical solution is given by Eq.(14) and
corresponds to an absolute minimum at the origin. On the other
hand, if $m_{R}^{2}<0$ Eq.(14) will correspond to a local
maximum at the origin while Eq.(\ref{21}) gives the position of
the two degenerate absolute minima. This solution describes a
broken symmetry state. It is also possible to obtain a
broken symmetry solution by using just the Eq.(\ref{21}). In
order to eq.(\ref{21}) to describe such a solution we must
demand that $<\rho>_{max}=0$, corresponding to a local
maximum at the origin, must be a solution to Eq.(\ref{21}). This
is possible only if $m_{R}^{2}=0$.
Thus, assuming $m_{R}^{2}=0$ and considering the
solution to Eq.(\ref{21}) $<\rho>_{min}=\sigma$, $\sigma$
corresponding to the minimum, we can solve Eq.(\ref{21}) for
$\lambda_{R}$:

\begin{equation}
\label{23}
\lambda_{R}=-\frac{9e^{4}}{8\pi^{2}}\ln\frac{e^{2}\sigma^{2}}
{\mu^{2}}.
\end{equation}
Substituting Eq.(\ref{23}) into Eq.(\ref{22}) with $m_{R}^{2}=0$
we have

\begin{equation}
\label{24}
V(\overline{\rho})=\frac{3e^{4}}{64\pi^{2}}\overline{\rho}^{4}\left(
\ln\frac{\overline{\rho}^{2}}{\sigma^{2}}-\frac{1}{2}\right).
\end{equation}
which is just the Coleman-Weinberg potential \cite{Coleman}. It is
important to note the non-perturbative character of our approach.
This
result was not obtained perturbatively by computing Feynman graphs.
In the usual loop expansion it is assumed that $\lambda_{R}$ is of
the same
order as $e^{4}$ and then terms proportional to $\lambda_{R}^{2}$ are
neglected. No such assumption is necessary here since our result
does not
depends upon perturbation theory. In perturbation theory this
assumption is
justified through renormalization group arguments \cite{Coleman}.
In our
approach this is not necesssary because from Eq.(\ref{22}) with
$m_{R}^{2}=0$
and Eq.(\ref{23}) the indenpendence of physical results on the
renormalization point is manifest. Thus, our results are obtained
in a
renormalization group invariant way.

We can give a mean field interpretation of our result. Mean field
theories
are generally characterized by a gap equation. In the Landau
approximation to $\phi^{4}$ theory the gap equation gives the
mean field critical indices of the Ising model. In our
case the gap equation is given by Eq.(\ref{15}). It is this gap
equation that, in the massless case, allows the elimination of the
arbitrary renormalization scale. Thus, we have obtained the
Coleman-Weinberg theory in a mean field like fashion. Note that we
achieve this result in a {\it physical} gauge, the unitary gauge.
 In the
Higgs mechanism this gauge exhibits the physical degrees of freedom
already at the tree level. We have just shown that the same is true
in
the case of the Coleman-Weinberg mechanism in an
Abelian theory. The Coleman-Weinberg
potential emerges as a "tree level" potential for the effective
action
Eq.(9).

Our derivation, being non-perturbative, is not restricted
to a small value of $e^{2}$. A large value of coupling constants
in the perturbative scheme is generally not allowed since this
may cause a breakdown of the perturbation series. We have
shown that the Coleman-Weinberg potential, originally a
perturbative result, can be established beyond the range of
validity of perturbation theory.
We can do a renormalization group analysis of the coupling
constant in this framework. By taking the fourth derivative at
$\sigma$ of the
potential, Eq.(\ref{22}), we obtain

\begin{equation}
\label{A}
V^{(4)}(\sigma)=\lambda_{R}+\frac{9e^{4}}{8\pi^{2}}\ln
\frac{e^{2}\sigma^{2}}{\mu^{2}}+\frac{33e^{4}}{8\pi^{2}}.
\end{equation}
We obtain that $V^{(4)}(\sigma)=\lambda_{R}$ if and only if we
have

\begin{equation}
\label{B}
\frac{33e^{4}}{8\pi^{2}}=-\frac{9e^{4}}{8\pi^{2}}\ln
\frac{e^{2}\sigma^{2}}{\mu^{2}}.
\end{equation}
From Eqs.(\ref{23}) and (\ref{B}) we obtain

\begin{equation}
\label{C}
\lambda_{R}=\frac{33e^{4}}{8\pi^{2}},
\end{equation}
which agrees with ref.\cite{Coleman}. Solving Eq.(\ref{B}) for
$e^{2}$ one obtains

\begin{equation}
\label{D}
e^{2}(\mu)=\frac{\mu^{2}}{\sigma^{2}}\exp\left(-
\frac{11}{3}\right).
\end{equation}
Thus, it does not matter at which energy scale we are working. We
can always choose to measure $e^{2}$ in units of
$\mu^{2}/\sigma^{2}$. Let us define $\alpha=e^{2}$ and
see what happens if we change the renormalization
scale in a such a way that $\mu{\longrightarrow}e^{t}\mu$.
From Eq.(\ref{D})
we obtain the following equation:

\begin{equation}
\label{E}
\frac{d\alpha}{dt}=2\alpha.
\end{equation}
Solving the above equation we get

\begin{equation}
\label{F}
\overline{\alpha}(t)=\alpha\exp(2t),
\end{equation}
which defines the running coupling constant $\overline{\alpha}(t)$.
It is easily seen that the theory is infrared free as
$t\longrightarrow-\infty$. Indeed, this is already manifest in
Eq.(\ref{D}). Also, in the ultraviolet end,
$t\longrightarrow\infty$, the running coupling constant
diverges. This is not problematic for us since we are not
performing a perturbative expansion. Also, we have already seen
that no matter how large is $\mu$, we can always choose to
measure the physical quantities in units of
$\mu^{2}/\sigma^{2}$. Note that our analysis does not coincide to
the usual one obtained from the conventional loop expansion
\cite{Coleman}. In the usual perturbative scheme it is found
that the running coupling diverges at a {\it finite} $t$,
which implies that a Landau ghost develops. However, it is
argued that this singularity arises in a scale which is
beyond the range of validity of the one loop result. Here
we established the one loop result non-perturbatively. Our
result implies that the running coupling diverges only for
infinite $t$ and there is no Landau ghost. This result has
implications concerning triviality questions. It has been
established rigorously that the $\phi^{4}$ theory is
trivial for $d{\geq}5$ \cite{Aizenman}.
It is almost certain that this is also true at $d=4$. It
has been speculated the impact of such a situation
(if it is true) in the Higgs models which involves a
scalar $\phi^{4}$ sector \cite{Callaway}.
In the non-Abelian Higgs model asymptotic freedom saves
the theory and the model is non-trivial although the
pure scalar sector appears to be trivial. However,
in the Abelian case there is no asymptotic freedom and
the ghost trouble occurs.
In the $\phi^{4}_{4}$
theory a Landau ghost seems to develop even at high orders in
perturbation theory. In our derivation of the Coleman-Weinberg
potential the self-coupling of the scalar field is related to
the coupling $e^{2}$ just as in the perturbative treatment.
However, the one loop renormalization group analysis shows that,
at least at this level, a Landau singularity arises. In the
present approach this does not happen. Thus, we conjecture that
in the case of the Coleman-Weinberg mechanism there is no
problem associated to the possible triviality of the pure
scalar theory. The coupling to the gauge field induces the
phenomenon of dimensional transmutation which protects
the theory against the triviality trouble.
This seems to happen also in the $\phi^{4}_{4}$ theory.
Recently, a
non-trivial phase to $\phi_{4}^{4}$ was found \cite{Langfeld}.
This phase is also associated to the phenomenon of
dimensional transmutation. This phenomenon does not occur in the
perturbative phase of $\phi^{4}_{4}$ \cite{Coleman}. However,
dimensional transmutation does
occur in a non-perturbative regime in the massless case. It is
found that the resulting expression for the effective
potential contains only one free parameter given by
$M_{0}=\mu^{2}\exp(96\pi^{2}/\lambda_{R})$ which has dimension
of mass squared \cite{Langfeld}. Therefore, dimensional
transmutation seems to be a possible path towards a non-trivial 
theory
in situatutions which there is no asymptotic freedom nor
non-trivial ultraviolet fixed points.

Let us consider now the finite temperature case. We can treat this
case by
standard methods \cite{Dolan}. At high temperature
we have

\begin{equation}
\label{25}
\lambda_{R}<\rho>_{T}^{2}=-6m_{R}^{2}-\frac{3e^{2}T^{2}}{2}+
\frac{9e^{3}T|<\rho>_{T}|}{2\pi}
-\frac{9e^{4}<\rho>_{T}^{2}}{8\pi^{2}}\ln\frac{e^{2}<\rho>_{T}
^{2}}{\mu^{2}},
\end{equation}
where we have assumed as before that $m_{R}^{2}=0$. Eq.(\ref{25})
is established by generalizing Eq.(\ref{15}) to finite
temperature using the prescription
$\int\frac{d^{4}p}{(2\pi)^{4}}{\longrightarrow}T\sum_{n}
\int\frac{d^{3}p}{(2\pi)^{3}}$, where the discrete sum is
over the Matsubara boson frequency $\nu_{n}=2n{\pi}T$
\cite{Dolan}. From Eqs.(\ref{23}) and (\ref{25}) we
obtain the following equation:

\begin{equation}
\label{26}
\frac{3e^{2}}{4\pi^{2}}<\rho>_{T}^{2}\ln\frac{
<\rho>_{T}^{2}}{\sigma^{2}}=\frac{3eT}{\pi}|<\rho>_{T}|
-T^{2}
\end{equation}

At high temperature we obtain the following expression for
the effective potential:

\begin{equation}
\label{27}
V(\overline{\rho})=\frac{3e^{4}}{64\pi^{2}}
\overline{\rho}^{4}
\left(\ln\frac{
\overline{\rho}^{2}}{
\sigma^{2}}-\frac{1}{2}\right)+\frac{e^{2}}{8}T^{2}
\overline{\rho}^{2}-\frac{e^{3}}{4\pi}T|\overline{\rho}|^{3}
\end{equation}
Note that from the above equation we can also obtain Eq.(\ref{26})
by imposing that $V'(<\rho>_{T})=0$.

In summary, we have obtained non-perturbatively the
effective potential for
massless scalar QED. We showed that with the treatment given here
the theory is free of Landau ghost singularity.
It is a question if we can treat the non-Abelian case similarly to
the
Abelian case. The
generalization of our approach to the non-Abelian case is not
straightforward because in a non-Abelian gauge theory the gauge
fields
interact and it is not possible to integrate out the
fields directly. Maybe this
problem can be solved partially by introducing auxiliary fields to
obtain a set of gap equations. This situation is under study.

This work was supported by Conselho Nacional de Desenvolvimento
Cientifico e Tecnologico - CNPq.

\end{document}